# Limits to Poisson's ratio in isotropic materials – general result for arbitrary deformation.

P. H. Mott and C. M. Roland

Chemistry Division, Naval Research Laboratory, Code 6120, Washington DC 20375-5342

*(April 3, 2012)*

**Abstract**

The lower bound usually cited for Poisson's ratio $\nu$ is −1, derived from the relationship between $\nu$ and the bulk and shear moduli. From consideration of the longitudinal and biaxial moduli, we recently determined that the lower bound on $\nu$ for isotropic materials is actually 1/5, a value also consistent with experimental measurements on real materials. Herein we generalize this result, first by analyzing expressions for $\nu$ in terms of six common elastic constants, and then by considering arbitrary strains. The results corroborate the prior finding that $1/5 \leq \nu$ for linear elasticity to be applicable.

## 1. Introduction

The ratio of lateral strain $\varepsilon_{22}$ to longitudinal strain $\varepsilon_{11}$ defines the elastic constant

$$\nu = -\frac{\varepsilon_{22}}{\varepsilon_{11}} \tag{1}$$

for a material under uniaxial stress $\sigma_{11}$. This constant is named for Poisson, who defined it in 1829 in developing his theory of linear elasticity, for which $\nu$ equaled ¼ for all solids [1]. Recent interest in auxetic materials ($\nu < 0$) [2,3] and nano-composites, in which Poisson's ratio is used to characterize mechanical behavior [4,5,6,7,8], has renewed attention to this quantity.

Much of the experimental investigations of the mechanical behavior of isotropic solids in the early 19th century were devoted to measuring $\nu$, to verify the single constant Poisson – Cauchy theory. Its disproof came sporadically: the first evidence appeared in 1848, when $\nu$ was found to be ca. ⅓ for five different oxide glasses and three different brasses [9]; and in 1859, when careful experiments found $\nu$ = 0.295 for steel [10]. Unfortunately, other less accurate measurements continued to provide support for the theory, and the controversy persisted into the 1860s. Lamé's two-constant linear elasticity theory for isotropic materials, introduced in 1852 [11], was adopted by most researchers soon thereafter in part because it accommodates variation in $\nu$ [12]. However, this did not prove it was correct.

According to classic elasticity, there are two unique elastic constants; to test a two-parameter theory requires measurement of three different constants. Elastic constants can be compared using Lamé's relation

$$\nu = \frac{1}{2} - \frac{E}{6B} \tag{2}$$

where $E$ and $B$ are the respective Young's and bulk moduli. There are two challenges to this approach: (1) the high precision required in the data (see review [13]); and (2) the complication that conventional solids are often non-linear even at strains as small as $10^{-5}$ [14,15]. Nonetheless since the error cause by the latter deviation is small, linear elasticity has been and remains widely applied in science and engineering. Experimental verification appeared in the early 1900s [16,17]. As seen in Fig. 1, measurements for iron, tin, aluminum, copper, silver, platinum, and lead [17] conform to Lamé's two-constant theory. With corroboration of the theory more than 100 years ago, common practice has become to limit characterizations of isotropic solid to determination of just two elastic constants, often obtained from shear and longitudinal wave speed measurements [18,19]; the other elastic constants are in turn calculated using the Lamé relations. This means, however, that the theory has been verified for conventional solids; that is, those in which $\nu \geq 0.2$.

As we recently showed [12], the conventionally accepted limits on Poisson's ratio are much lower than the experimental range in Fig. 1. The theoretical limits are found from

$$G = B\frac{3(1-2\nu)}{2(1+\nu)} \tag{3}$$

where $B$ and $G$ are the respective bulk and shear moduli. This relation describes deformation in terms of respective changes in size and shape. In order to minimize strain energy at equilibrium to avoid spontaneous deformation, $G$ and $B$ must be positive, leading to the oft-stated "thermodynamically admissible" range [12,20]

$$-1 < \nu < \tfrac{1}{2} \tag{4}$$

A recent analysis [21] reaffirmed this range when consideration is limited to elastic constants that can be expressed as linear combinations of the two Lamé constants, namely *B* and the longitudinal modulus (defined below).

Experimental $\nu$ for isotropic materials occupy a much narrower range. An examination of literature data encompassing more than 3,000 measurements on 596 different substances over a wide range of temperature and pressure, including pure elements, engineering alloys, polymers, ceramics, and glasses, revealed that with very few exceptions (e.g., porous quartz or very hard materials such as diamond and beryllium), $\nu \geq 0.2$ for isotropic, homogeneous materials [22,23]. Thus, the lower limit in eq. 4 does not represent the behavior of real materials.

Notwithstanding its conceptual appeal, there is no mathematical or physical justification for preferring *G* and *B* over other pairs of constants for determination of the limits on Poisson's ratio. For example, we have shown from the roots of a quadratic expression for $\nu$ that the range in eq. 4 is split into [23]

$$\begin{aligned} -1 < \nu \leq 1/5 \\ 1/5 \leq \nu < 1/2 \end{aligned} \qquad (5)$$

Since elastic properties are unique, only one range can be valid. Moreover, the lower limit of 1/5 agrees with experimental data. Thus, the more restrictive upper range, $1/5 \leq \nu < 1/2$, is the correct one. The argument might be made that the range extending to $-1 < \nu$ is mathematically valid, and hence represents an acceptable bound. However, rejection of spurious roots is not uncommon when an analysis produces two or more solutions; physical considerations are applied to eliminate roots that are false. Examples include the Landau-Lifshitz equation for the motion of a charge [24], analysis of projectile trajectories in air [25], Pythagoras' theorem for right triangles, and more generally in the solutions of ordinary differential equations [26].

This revised, more restrictive lower bound on Poisson's ratio is important because it means that whenever a material has $\nu < 0.2$, the equations of linear elasticity do not apply. In this work we first extend the analysis of ref. [23] to all commonly defined elastic constants, in order to obtain their associated limits for Poisson's ratio. We then generalize these results to arbitrary deformation mode. Our previous conclusion [23], that the minimum of $\nu$ for an isotropic material is 1/5, is shown to be general for materials for which the equations of classic elasticity are valid.

## 2. Limits on *ν* from common elastic constants

For an isotropic solid with strain tensor $\varepsilon_{ij}$, the reversible work of deformation is [12]

$$2W = (\lambda + 2\mu)(\varepsilon_{11} + \varepsilon_{22} + \varepsilon_{33})^2 + \mu(\varepsilon_{11}{}^2 + \varepsilon_{22}{}^2 + \varepsilon_{33}{}^2 - 4\varepsilon_{22}\varepsilon_{33} - 4\varepsilon_{33}\varepsilon_{11} - 4\varepsilon_{11}\varepsilon_{22}) \quad (6)$$

where $\lambda$ and $\mu$ (=$G$) are the Lamé constants. Differentiation with respect to $\varepsilon_{ij}$ defines the stress tensors $\sigma_{ij}$. When uniaxial loading is substituted (i.e., $\sigma = \sigma_{11}$ and all other $\sigma_{ij} = 0$),

$$E = \frac{\mu(3\lambda + 2\mu)}{\lambda + \mu}, \quad \nu = \frac{\lambda}{2\lambda + 2\mu} \quad (7)$$

This procedure can be carried out for any deformation or loading geometry to define the corresponding stiffness [27]. These definitions are combined to obtain relations between the elastic constants. For example, for longitudinal loading ($\varepsilon = \varepsilon_{11}$ and all other $\varepsilon_{ij} = 0$) we obtain

$$M = \frac{1-\nu}{(1-2\nu)(1+\nu)} E \quad (8)$$

where $M$ is the longitudinal modulus.

Table 1 lists all of the equations for Poisson's ratio from commonly defined moduli. Included are expressions that involve the biaxial stress modulus $H$, defined when $\sigma = \sigma_{11} = \sigma_{22}$ and all other $\sigma_{ij} = 0$, and the biaxial strain modulus $I$, defined when $\varepsilon = \varepsilon_{11} = \varepsilon_{22}$ and all other $\varepsilon_{ij} = 0$. $I$ is unusual, but is included here as the counterpart to $H$. The second column in the Table shows the restrictions on $\nu$ arising from the requirement that all elastic moduli are greater than zero. It is seen that the conventional limits, $-1 < \nu < ½$, follow from eqs. T1 and T2. The other linear expressions lead to wider ranges for $\nu$. Of course, the more restrictive limits for Poisson's ratio is the governing range, since all broader ranges are also satisfied.

Of special interest are the four quadratic relations, eqs. T12 – T15. These arise from stress-strain counterparts, such as $E$ (defined from a stress) and $M$ (defined from a strain). Note that if $E/M$ is substituted for $H/I$, eq. T13 becomes eq. T12, and therefore the two equations are identical; i.e.,

$$\frac{E}{M} = \frac{H}{I} \quad (9)$$

Each quadratic relation in Table 1 has two roots that limit the span of Poisson's ratio. These relations are plotted in Fig. 2, with the positive roots denoted by a solid line and the negative with a dashed line. The roots converge at smoothly continuous maxima, all giving the same bounds of −1 and ½. Restricting $v$ to real numbers means that:

1. Eqs. T12 and T13: $0 < E/M \leq 1$ with the same range for $H/I$. The two roots of this expression have ranges $-1 < v \leq 0$ and $0 \leq v < ½$. This equation also produces real values if $E/M \geq 9$, which has two roots with ranges $1 < v \leq 2$ and $2 \leq v < \infty$; however, this solution is discarded because it falls beyond the bounds of eq. 4.
2. Eq. T14: $0 < E/I \leq 9/8$; the two roots have the ranges $-1 < v \leq -¼$ and $-¼ \leq v < ½$.
3. Eq. T15: $0 < H/M \leq 9/8$; the two roots have the ranges $-1 < v \leq 1/5$ or $1/5 \leq v < ½$.

There are companion relations for $G$ and $B$, and these quadratic equations have interconnected roots. For example, the counterpart to eq. T15 for the bulk modulus is

$$B = \frac{M}{6}\left[3 \pm \left(9 - 8\frac{H}{M}\right)^{½}\right] \tag{10}$$

and, having the same argument for the square root as in eq. T15, restricts $0 < H/M \leq 9/8$ for this expression to be real. The negative root has the range $0 < B/M \leq ½$, $½ \leq B/M < 1$ for the positive root. It can be shown that the positive root is linked to the positive root of eq. T15 and vice-versa; that is, if $½ \leq B/M < 1$, then $1/5 \leq v < ½$.

Quadratic expressions with two possible solutions for $G$, $B$, and $v$ are at odds with the behavior of real materials, which have unique elastic constants for any thermodynamic state. Therefore, only one set of solutions can be valid.

## 3. Limits on *v* for arbitrary deformations

The considered elastic constants – shear $G$, hydrostatic pressure or dilatation $B$, uniaxial stress $E$, uniaxial strain $M$, biaxial stress $H$, and biaxial strain $I$ – permute a single stress or strain through the available tensor combinations for an isotropic material. However, the possibility exists that more restrictive limits on $v$ can be found from other elastic constants derived from more complex combinations of stress or strain. To examine this, we introduce two, continuously variable elastic constants. The first is a biaxial stress with $\sigma_{11} = \sigma$ and $\sigma_{22} = y\sigma$, where $y$ is a constant describing the fraction of biaxial stress, $0 \leq y \leq 1$; all other $\sigma_{ij} = 0$. The elastic constant for this variable stress geometry is

$$H_y = \frac{E}{1 - yv} \tag{11}$$

When $y = 0$ (uniaxial loading), $H_0 = E$; when $y = 1$ (biaxial stress), eq. 11 becomes eq. T8.

For the second constant, consider a variable biaxial strain $\varepsilon_{11} = \varepsilon$; $\varepsilon_{22} = \beta\varepsilon$, where $\beta$ is the fraction of biaxial strain, $0 \leq \beta \leq 1$; and all other $\varepsilon_{ij} = 0$. The elastic constant for this variable strain geometry is

$$I_\beta = \frac{1-\nu\,(1-\beta)}{1-\nu} M \tag{12}$$

Similarly, when $\beta = 0$, $I_0 = M$ (longitudinal deformation), and when $\beta = 1$, eq. 12 becomes eq. T9, corresponding to biaxial strain. These expressions define the elastic stiffness for any mixture of one or two dimensional stress or strain.

From the equations in Table 1, other relations that involve $H_y$ and $I_\beta$ can be derived. Of particular interest is

$$\nu = \frac{I_\beta}{4I_\beta + 2y(1-\beta)H_y} \left\{ (1-\beta+y)\frac{H_y}{I_\beta} - 1 \pm \left[ 9-(10-2\beta-2y+4\beta y)\frac{H_y}{I_\beta} + (1-\beta-y)^2 \frac{H_y{}^2}{I_\beta{}^2} \right]^{1/2} \right\} \tag{13}$$

This equation combines the four quadratic expressions for Poisson's ratio into a single, continuous function. Each of the four quadratic expressions for $\nu$ in Table 1 can be recovered by substituting the respective values for $y$ and $\beta$. Intermediate values $y$ and $\beta$ produce curves that lie between these extremes. Shown in Fig. 2 is the curve for $y = ½$ and $\beta = 0$, which falls between the $H/M$ and $E/M$ curves. Likewise, the two roots of eq. 13 meet without discontinuity. This common point is defined as $\nu^*(y,\beta)$ at $H_y^*/I_\beta^*$; it divides Poisson's ratio into the ranges $-1 < \nu \leq \nu^*$ and $\nu^* \leq \nu < ½$. Since the upper span corresponds to experimental data [28,29], it is of interest to determine the lower limit $\nu^*$. This point is found when the two roots are equal, which occurs when

$$9-(10-2\beta-2y+4\beta y)\frac{H_y^*}{I_\beta^*} + (1-\beta-y)^2\left(\frac{H_y^*}{I_\beta^*}\right)^2 = 0 \tag{14}$$

This expression has the solutions

$$\frac{H_y^*}{I_\beta^*} = \frac{5-y-\beta+2\beta y \pm 2[(\beta^2-\beta-2)(y^2-y-2)]^{1/2}}{(1-\beta-y)^2} \tag{15}$$

The positive root is rejected because it returns $H_y^*/I_\beta^* \geq 9$, producing $\nu > 1$, which is beyond the bounds from eq. 4. Note this corresponds to $E/M \geq 9$ , which was discarded in Section 2 above.

The correct root of eq. 15 is plotted in Fig. 3, showing $H_y^*/I_\beta^*$ (top) and $v^*$ (bottom) for values of $y$ and $\beta$. $H_y^*/I_\beta^*$ has the range $1 < H_y^*/I_\beta^* \leq 9/8$ and increases symmetrically from $y = \beta$, where $H_y^*/I_\beta^* = 1$. Values of $v^*$ have the range $-1/4 < v^* \leq 1/5$, varying anti-symmetrically about $v^* = 0$, where $y = \beta$. The four corners of the figure, where respectively $v^*$ = (i) 1 at $\beta = 0$, $y = 0$; (ii) 1/5 at $\beta = 0$, $y = 1$; (iii) −1/4 at $\beta = 1$, $y = 0$; and (iv) 0 at $\beta = 1$, $y = 1$, correspond to $v^*(E,M)$, $v^*(H,M)$, $v^*(E,I)$, and $v^*(H,I)$. Thus, the ranges of $v$ for specific conditions of stress and strain (Fig. 1) are merged into a single continuous function describing arbitrary stress and strain. Fractional values of $y$ and $\beta$ in eq. 13 determine $v^*$ for any combination of two-dimensional stress or strain. Again, the most restrictive range is the correct range because it accommodates the other ranges; thus, the lower bound for $v$ is 1/5 for any stress and strain.

Note that eq. 15 is undefined when $\beta + y = 1$. For this condition, the solution for $H_y^*/I_\beta^*$ is found by substituting $a - y = \beta$ and taking the limit $a \to 1$ by twice applying L'Hôpital's rule. The result is

$$H_y^*/I_\beta^* = 1 - \frac{(1-2y)^2}{4(y^2 - y - 2)} \tag{16}$$

This demonstrates that there is no discontinuity when $\beta + y = 1$.

The companion quadratic relations for $G$ and $B$ are

$$G = \frac{I_\beta}{4-8\beta}\left\{3 + (1-\beta+y)\frac{H_y}{I_\beta} \mp \left[9 - (10 - 2\beta - 2y + 4\beta y)\frac{H_y}{I_\beta} + (1-\beta-y)^2\frac{{H_y}^2}{{I_\beta}^2}\right]^{1/2}\right\} \tag{17}$$

$$B = \frac{I_\beta}{6+6\beta}\left\{3 - (1-\beta+y)\frac{H_y}{I_\beta} \pm \left[9 - (10 - 2\beta - 2y + 4\beta y)\frac{H_y}{I_\beta} + (1-\beta-y)^2\frac{{H_y}^2}{{I_\beta}^2}\right]^{1/2}\right\} \tag{18}$$

The inverted ± sign in eq. 17 denotes that its negative root is linked to the positive roots of eqs. 13 and 18.

## 4. **Exceptions**

As stated in the introduction, isotropic materials exist for which $v < 1/5$, although they are rare. Homogenous materials which show this behavior include pyrite [30], α-cristobalite [31], diamond [32,33,34], a $TiNb_{24}Zr_4Sn_{7.9}$ (β-type titanium) alloy [35], boron nitride [36], α-beryllium [37], and certain silicate glasses [38]. In the former cases (pyrite, cristobalite, diamond), elastic properties have been determined from vibrational measurements of single crystals, and aggregate isotropic behavior is inferred. For the titanium alloy, boron nitride, beryllium, and $SiO_2$ glasses, elastic properties of the aggregate were determined by vibrational methods, in which two elastic constants are measured, with Poisson's ratio in turn found from the expressions in table 1. Thus, the Lamé relations have not been tested for homogeneous solids in which $\nu < 1/5$.

There are recent reports of auxetic behavior in crystalline materials that exhibit negative $\nu$ only in certain directions [39,40]. However when the aggregate isotropic behavior is examined, these substances show the conventional behavior, $\nu \geq 1/5$. There are also a class of heterogeneous foams that have negative Poisson's ratio [3]. These auxetic foams exhibit non-linear mechanical properties [41], so that the application of linear elasticity is problematic. Efforts have been made to fit the behavior to more complicated elasticity models, with limited success [42]. Recent investigations of larger scale, two-dimensional skeletal structures, both experimental [43,44] and theoretical [45], also discovered interesting auxetic behavior, but again linear elasticity does not apply to deformations larger than mathematically infinitesimal.

## 5. Summary

The equations of classic elasticity impose restrictions on the values of Poisson's ratio. Any pair of elastic constants leads to various expressions for the bounds on $\nu$; however, for mutual consistency the most restrictive limits are the correct ones. The result, $1/5 \leq \nu < ½$, is shown to be the valid range for any isotropic material subjected to arbitrary loading or deformation. This range comports with the values of $\nu$ for the vast majority of isotropic materials. For those materials showing deviations of $\nu$ from these limits, the equations of classic elasticity cannot be applied.

**Acknowledgements.**

We acknowledge useful discussions with Prof. R. S. Lakes and many thoughtful remarks by Dr. D.M. Fragiadakis. This work was supported by the Office of Naval Research.

## References

1 Poisson, S.D., 1829. Mémoire sur l'équilibre et le movement des corps élastiques. Mém. de l'Acad. Sci. 8, 357.

2 Liu, Y., Hu, H., 2010. A review on auxetic structures and polymeric materials. Sci. Res. Essays 5, 1052-1063

3 Lakes, R., 1987. Foam structures with a negative Poisson's ratio. Science 235, 1038-1040.

4 Wang, C.M., Tay, Z.Y., Chowdhuary, A.N.R., Duan, W.H., Zhang, Y.Y. Silvestre, N., 2011. Examination of cylindrical shell theories for buckling of carbon nanotubes. Int. J. Struc. Stab. Dyn. 11, 1035-1058.

5 Chen, L., Liu, C., Wang, J., Zhang, W., Hu, C., Fan, S., 2009. Auxetic materials with large negative Poisson's ratios based on highly oriented carbon nanotube structures. Appl. Phys. Let. 94, 253111.

6 Coluci, V.R., Hall, L.J., Kozlov, M.E., Zhang, M., Dantas, S.O., Galvao, D.S., Baughman, R.H., 2008. Phys. Rev. B 78, 115408.

7 Lee, D., Wei, X.D., Kysar, J.W., Hone, J., 2008. Measurement of the elastic properties and intrinsic strength of monolayer graphene. Science 321, 385-388.

8 Robertson, C.G., Bogoslovov, R., Roland, C.M. 2007. Effect of structural arrest on Poisson's ratio in nanoreinforced elastomers. Phys. Rev. E 75, 051403.

9 Wertheim, G., 1848. Mémoire sur l'élasticité des corps solides homogènes. Annales of de Chemie et de Physique, 3rd series 23, 52-95; cited by Bell, J. F. 1973. The experimental foundations of solid mechanics. in Handbuch der Physik 1973. volume VIa/1, ed. C. Truesdale.

10 Kirchhoff, G., 1859. Pogg. Ann. Phys. Chem. 108, 316.

11 Lamé, G. 1852. Leçons sur la théorie mathématique de l'élasticité des corps solides, Bachelier, Paris.

12 Love, A.E.H. 1944. A Treatise on the Mathematical Theory of Elasticity, Dover, New York; 1966. Mechanical Behavior of Materials, F.A. McClintock and A.S. Argon, eds., Addison-Wesley, Reading, Massachusetts.

13 Bell, J.F., 1973. The experimental foundations of solid mechanics. in Handbuch der Physik volume VIa/1, ed. C. Truesdale.

14 Hartman. W.F., 1967. The applicability of the generalized parabolic deformation law to a binary alloy. PhD dissertation, Johns Hopkins University.

15 Bell, J.F. 1968. The physics of large deformation of crystalline solids. Springer Tracts in Natural Philosophy, vol 14 Berlin: Springer.

16 Grüneisen. E.A., 1908. Torsionmodul, verhältnis von querkontraktion zu Längsdilatation und kubische kompressibiltät. Ann. Phys. (Berlin) 330, 825-851.

17 Grüneisen, E.A., 1910. Einfluß der temperature auf die kompressibilität der metalle. Ann. Phys. (Berlin) 338, 1239-1274.

18 Bradfield, G. 1964. Use in Industry of Elasticity Measurements in Metals with the help of Mechanical Vibrations, Her Majesty's Stationary Office, London.

19 Spinner, S., 1954. Elastic moduli of glasses by a dynamic method," J. Am. Ceram. Soc. 37, 229-234.

20 Greaves, G.N., Greer, A.L., Lakes, R.S., Rouxel, T., 2011. Poisson's ratio and modern materials. Nature Matl. 10, 823-837.

21 Tarumi, R., Ledbetter, H., Shibutani, Y., 2012. Some remarks on the range of Poisson's ratio in isotropic linear elasticity. Phil. Mag. 92, 1287-1299.

22 Simmons, G., Wang, H. 1971. Single Crystal Elastic Constants and Calculated Aggregate Properties: A Handbook, MIT press, Cambridge Massachusetts.

23 Mott, P.H., Roland, C.M. 2009. Limits to Poisson's ratio in isotropic materials. Phys. Rev. B 80, 132104.

24 Mares, R., Ramírez-Baca, P.I. Ares de Parga, G. 2010. Lorentz-Dirac and Landau-Lifshitz equations without mass renormalization: ansatz of Pauli and renormalization of the force. J. Vectorial Relativity 5, 1-8.

25 Warburton, R.D.H., Wang, J,. Burgdörfer, J., 2010. "Analytic approximations of projectile motion with quadratic air resistance. J. Service Science and Management 3, 98-105.

26 Shampine, L.F., Thompson, S., Kierzenka, J.A., Byrne, G.D., 2005. Appl. Math. Comp. 170, 556-559.

27 Tschoegl, N.W., Knauss, W.G., Emri, I. 2002. "Poisson's ratio in linear viscoelasticity: a critical review," Mech. Time-Depend. Mater. 6, 3-51.

28 Mott, P.H., Dorgan, J.R., Roland, C.M., 2008. The bulk modulus and Poisson's ratio of "incompressible" materials. J. Sound Vibr. 312, 572-575.

29 Mott, P.H., Roland, C.M., 2010. Response to 'Comment on paper 'The bulk modulus and Poisson's ratio of "incompressible" materials', J. Sound Vibr. 329, 368-369.

30 Simmons, G., Birch, F,. 1963. Elastic constants of pyrite. J. Appl. Phys. 34, 2736-2738.

31 Yeganeh-Haeri, A., Weidner, D.J., Parise, J.B., 1992. Elasticity of α-cristobalite: a silicon dioxide with a negative Poisson's ratio. Science 257, 650-652.

32 Anastassakis, E., Siakavellas, M., 2001. Elastic properties of textured diamond and silicon. J. Appl. Phys. 90, 144-152.

33 D'Evelyn, M.P., Zgonc, K., 1997. Elastic properties of polycrystalline cubic boron nitride and diamond by dynamic resonance method. Diamond Relat. Mater. 6, 812-816.

34 Klein, C.A., Cardinale G.F., 1993. Young's modulus and Poisson's ratio of CVD diamond. Diamond Relat. Mater. 2, 918-923.

35 Hao, Y.L., Li, S.J., Sun, B.B., Sui, M.L., Yang, R., 2007. Ductile titanium alloy with low Poisson's ratio. Phys. Rev. Lett. 98, 216405.

36 D'Evelyn, M.P., Zgonc, K., 1997. Elastic properties of polycrystalline cubic boron nitride and diamond by dynamic resonance method. Diamond Relat. Mater. 6, 812-816.

37 Migliori, A., Ledbetter, H., Thoma, D.J., Darling, T.W., 2004. Beryllium's monocrystal and polycrystal elastic constants. J. Appl. Phys. 95, 2436-2440.

38 Rouxel, T., Ji, H., Hammouda, T., Moréac, A., 2008. Poisson's Ratio and the densification of glass under high pressure. Phys. Rev. Lett. 100, 225501.

39 Baughman, R.H., Shacklette, J. M., Zakhidov, A.A., Stafström, S., 1998. Negative Poisson's ratio as a common feature of cubic metals. Nature 392, 362-364.

40 Rovati, M. 2004. "Directions of auxeticity for monoclinic crystals," Scripta Mater. 51, 1087-1091.

41 Choi, J.B., Lakes, R.S. 1992. "Non-linear properties of polymer cellular materials with a negative Poisson's ratio," J. Mater. Sci. 27, 4678-4684.

42 Anderson, W.B., Lakes, R.S., 1994. Size effects due to Cosserat elasticity and the surface damage in closed-cell polymethacrylimide foam. J. Mater. Sci. 29, 6413-6419.

43 Fozdar, D.F., Soman, P., Lee, J.W., Han, L-H., Chen, S. 2011. Three-dimensional polymer constructs exhibit a tunable negative Poisson's ratio. Adv. Mater. 21, 2712-2720.

44 Mitschke, H., Schwerdtfeger, J., Schury, F., Stingl, M., Körner, C., Singer, R.F., Robins, V., Mecke, K., Schröder-Turk, G.E. 2011. Finding auxtic frameworks in periodic tessellations, Adv. Mater. 23, 2669-2674.

45 Milton, G.W. 1992. Composite materials with Poisson's ratios close to −1. J. Mech. Phys. Solids 40, 1105-1137.

Table 1: Relations between elastic constants that include Poisson's ratio

| RELATION | (eq.) | RESTRICTIONS ON $\nu$ |
|---|---|---|
| $\nu = \frac{3B-2G}{6B+2G}$ | (T1) | $-1 < \nu < \frac{1}{2}$ |
| $\nu = \frac{H-3B}{H-6B}$ | (T2) | $-1 < \nu < \frac{1}{2}$ |
| $\nu = \frac{3B-M}{3B+M}$ | (T3) | $-1 < \nu < 1$ |
| $\nu = \frac{H-2G}{H+2G}$ | (T4) | $-1 < \nu < 1$ |
| $\nu = \frac{M-2G}{2M-2G}$ | (T5) | $-\infty < \nu < \frac{1}{2}$ |
| $\nu = \frac{1}{2} - \frac{E}{6B}$ | (T6) | $-\infty < \nu < \frac{1}{2}$ |
| $\nu = \frac{1}{2} - \frac{G}{I}$ | (T7) | $-\infty < \nu < \frac{1}{2}$ |
| $\nu = 1 - \frac{E}{H}$ | (T8) | $-\infty < \nu < 1$ |
| $\nu = 1 - \frac{M}{I}$ | (T9) | $-\infty < \nu < 1$ |
| $\nu = \frac{E}{2G} - 1$ | (T10) | $-1 < \nu < \infty$ |
| $\nu = \frac{3B}{I} - 1$ | (T11) | $-1 < \nu < \infty$ |
| $\nu = \frac{1}{4}\left[\frac{E}{M} - 1 \pm \left(\frac{E^2}{M^2} - 10\frac{E}{M} + 9\right)^{\frac{1}{2}}\right]$ | (T12) | $0 < \frac{E}{M} \le 1$: $-1 < \nu < 0$ *or* $0 < \nu < \frac{1}{2}$ |
| $\nu = \frac{1}{4}\left[\frac{H}{I} - 1 \pm \left(\frac{H^2}{I^2} - 10\frac{H}{I} + 9\right)^{\frac{1}{2}}\right]$ | (T13) | $0 < \frac{H}{I} \le 1$: $-1 < \nu < 0$ *or* $0 < \nu < \frac{1}{2}$ |
| $\nu = -\frac{1}{4}\left[1 \mp \left(9 - 8\frac{E}{I}\right)^{\frac{1}{2}}\right]$ | (T14) | $0 < \frac{E}{I} \le \frac{9}{8}$: $-1 < \nu \le -\frac{1}{4}$ *or* $-\frac{1}{4} \le \nu < \frac{1}{2}$ |
| $\nu = \frac{M}{2H+4M}\left[2\frac{H}{M} - 1 \pm \left(9 - 8\frac{H}{M}\right)^{\frac{1}{2}}\right]$ | (T15) | $0 < \frac{H}{M} \le \frac{9}{8}$: $-1 < \nu \le \frac{1}{5}$ *or* $\frac{1}{5} \le \nu < \frac{1}{2}$ |

**Figure Captions**

**Figure 1.** Experimental data from Grüneisen [17], demonstrating the validity of Lamé's quadratic theory of linear elasticity.

**Figure 2.** Poisson's ratio as a function of the ratio of the indicated elastic constants, with positive roots shown indicated by the solid lines and negative roots with dashed lines. Also included are the two roots of eq. 13 with $y = ½$ and $\beta = 0$. The limits encompassing all moduli is $1/5 \leq \nu < ½$.

**Figure 3.** (Top) $H_y^*/I_\beta^*$ obtained from the negative roots of eq. 15 for all values of $y$ and $\beta$, and the line $y + \beta = 1$ from eq. 16. (Bottom) Value of Poisson's ratio $\nu^*$ satisfying both roots of eq.13, which corresponds to the minimum value of the range $\nu^* \leq \nu < ½$. The maximum of $\nu^*$ in the figure, 1/5, is the governing value.

FIG 1

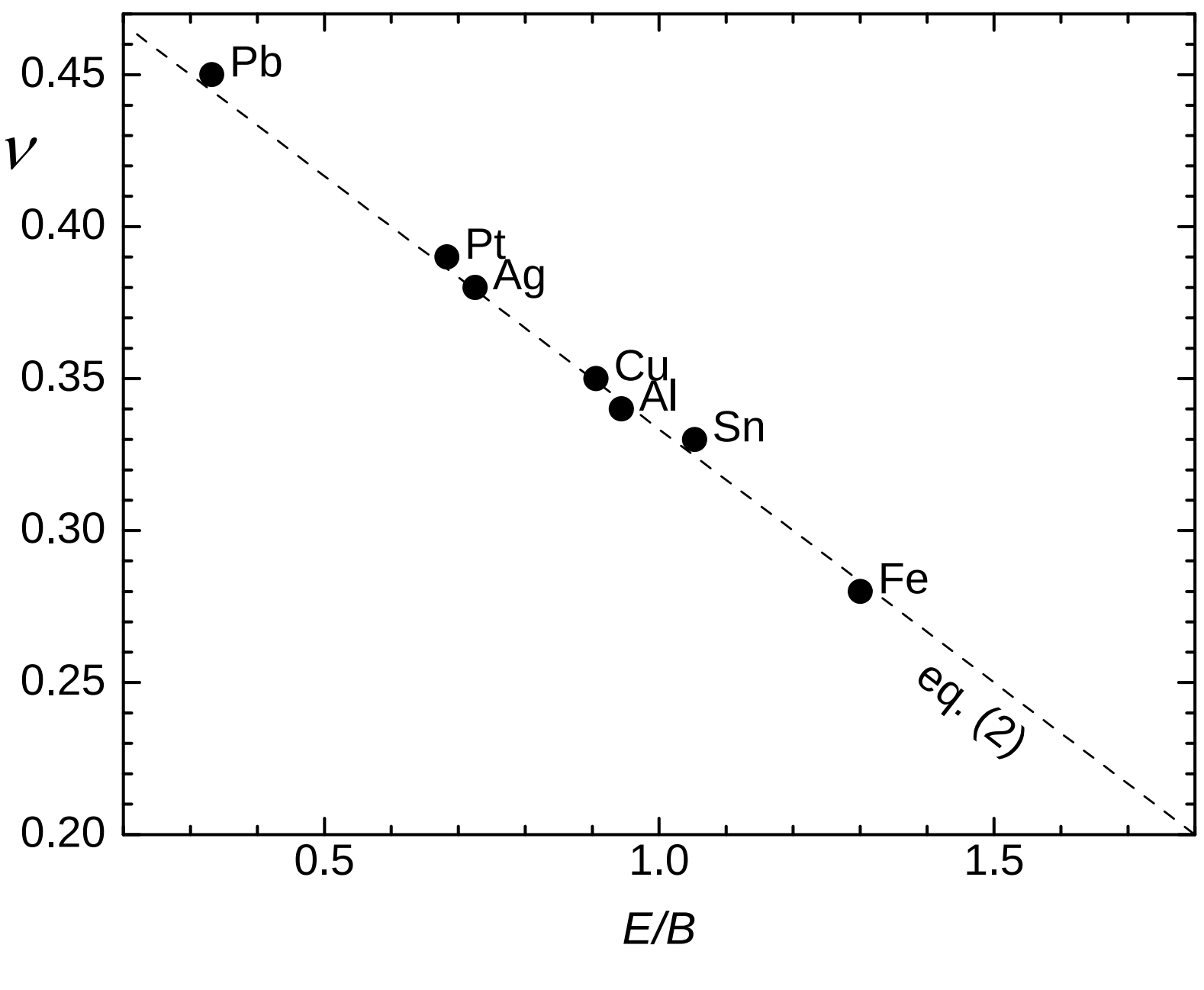

FIG 2

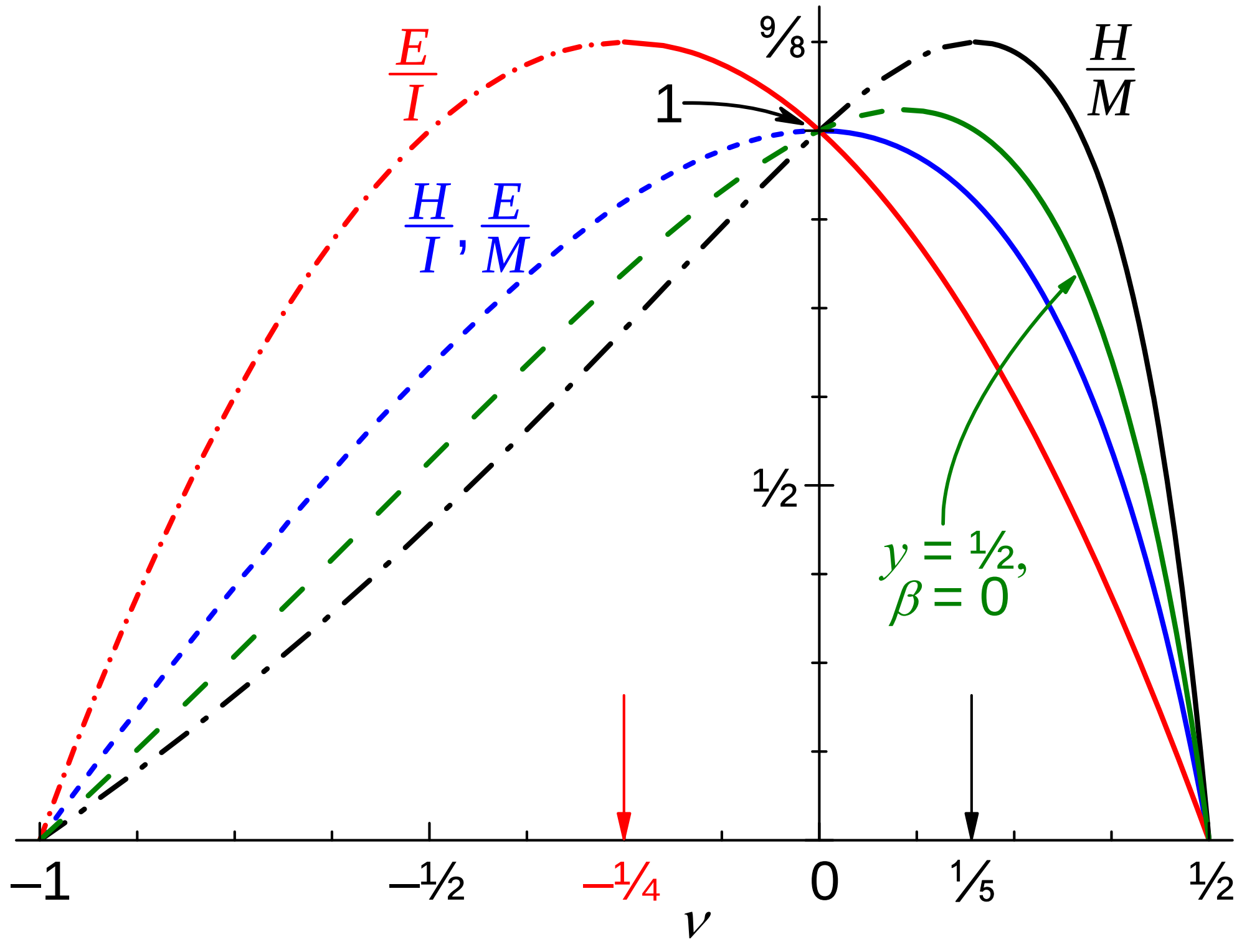

FIG 3

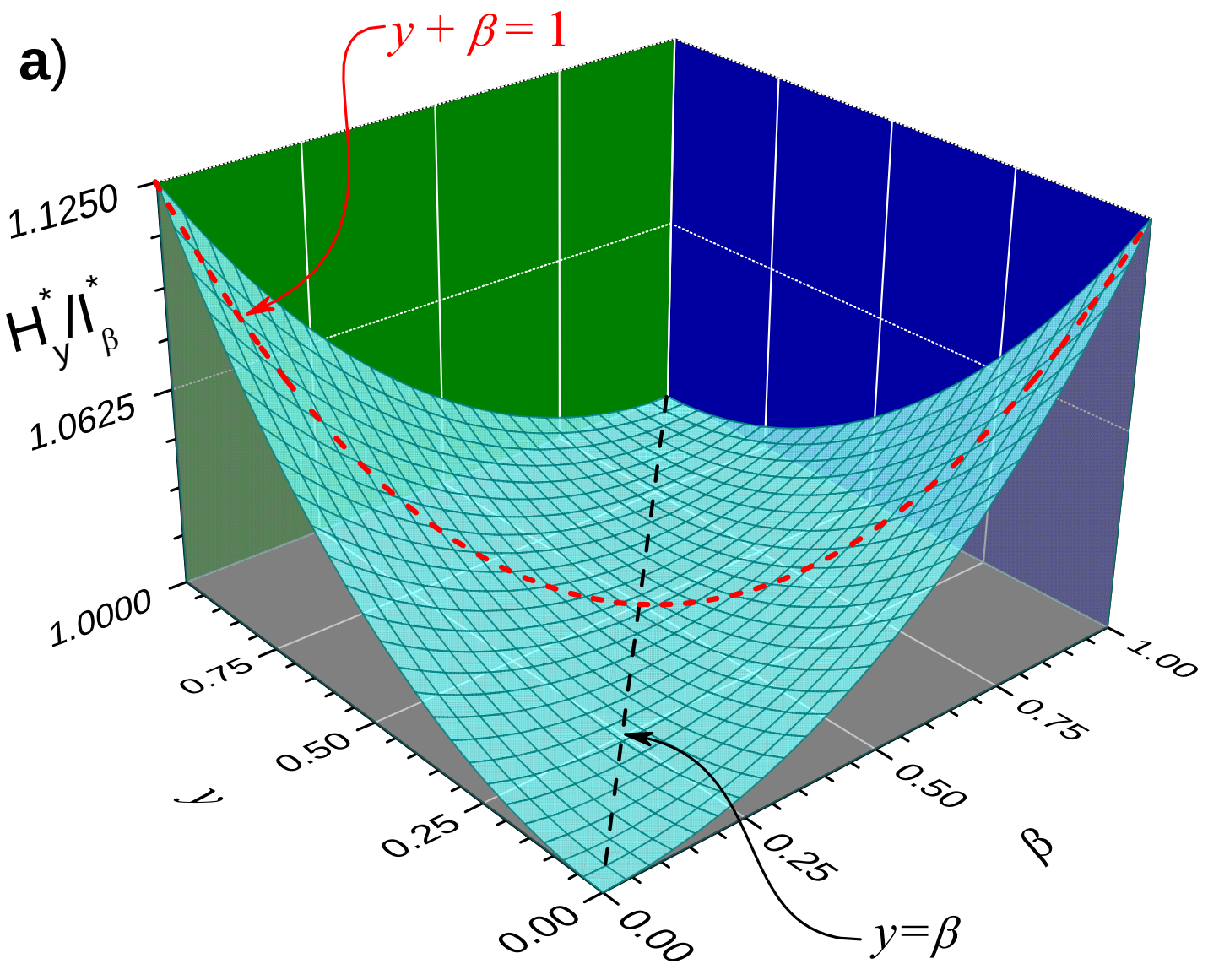

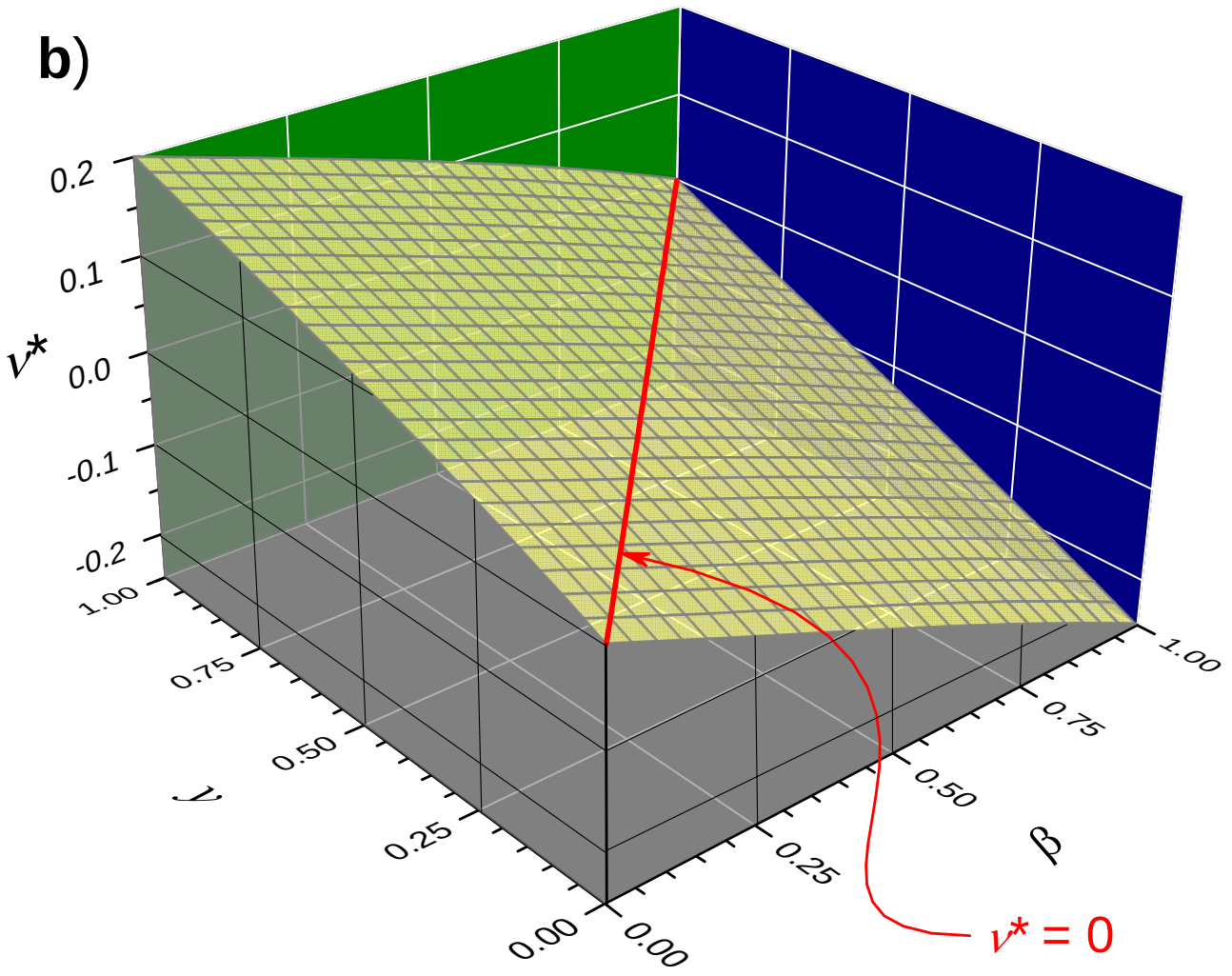